\pdfoutput=1

\documentclass[11pt]{article}

\usepackage{ACL2023}

\usepackage{times}
\usepackage{latexsym}
\usepackage{tikz}

\usepackage[T1]{fontenc}

\usepackage[utf8]{inputenc}

\usepackage{microtype}

\usepackage{inconsolata}

\title{ivrit.ai: A Comprehensive Dataset of Hebrew Speech for AI Research and Development}

\author{
  Yanir Marmor \\
  Weizmann Institute of Science \\
  \small \texttt{yanirmarmor@weizmann.ac.il}
  \And
  Kinneret Misgav \\
  Hadassah Research Fund \\
  \small \texttt{mkinneret@hadassah.org.il}
  \And
    Yair Lifshitz \\
    \\
    \small \texttt{yair@lifshitz.io}
}

\begin{document}

\newcommand{\totalHours}{3,300 }
\newcommand{\totalWords}{30 million }

\maketitle
\begin{abstract}

We introduce \textit{ivrit.ai}, a comprehensive Hebrew speech dataset, addressing the distinct lack of extensive, high-quality resources for advancing Automated Speech Recognition (ASR) technology in Hebrew. With over \totalHours speech hours and a over a thousand diverse speakers, \textit{ivrit.ai} offers a substantial compilation of Hebrew speech across various contexts. It is delivered in three forms to cater to varying research needs: raw unprocessed audio; data post-Voice Activity Detection, and partially transcribed data. The dataset stands out for its legal accessibility, permitting use at no cost, thereby serving as a crucial resource for researchers, developers, and commercial entities. \textit{ivrit.ai} opens up numerous applications, offering vast potential to enhance AI capabilities in Hebrew. Future efforts aim to expand \textit{ivrit.ai} further, thereby advancing Hebrew's standing in AI research and technology.
\end{abstract}

\section{Introduction}
Automated Speech Recognition (ASR; also known as speech-to-text) technology holds vast potential for enhancing various processes involving human speech. Nevertheless, its effectiveness is not uniformly distributed across languages. While some languages significantly benefit from ASR tools, others, such as Hebrew, find these technologies to be underwhelming. This often results in the continued use of human scribes when transcriptions are not available.

The initial phase of creating precise and efficient ASR tools requires an extensive corpus of high-quality speech data. To our knowledge, such a dataset for Hebrew has not been publicly available until now. We introduce \textit{ivrit.ai} ("Ivrit" is how Hebrew speakers pronounce the name of their language), a comprehensive Hebrew speech dataset designed for AI research and development. \textit{ivrit.ai} is designed to empower advanced AI technologies to function smoothly in Hebrew, enabling AI to read, write, listen to, and articulate the language with fluency. We view this as a strategic initiative to provide Hebrew speakers and their communities with access to superior AI technology in a practical manner under a favorable license, thus enabling interested commercial parties to use this data at no cost. \textit{ivrit.ai} contains approximately \totalHours hours of Hebrew speech, collected from a diverse range of online platforms including podcasts and other audio content. Given the wide variety of speech types and topics within \textit{ivrit.ai}, it serves as a valuable resource for advancing AI in Hebrew.

\subsection{Automated Speech Recognition}

Oral communication is a crucial component of human interaction. It communicates words and meaning through intonation, pitch (i.e., how high or low the voice sounds), pace, and subtle non-verbal cues, together forming the rich tapestry of human conversation. Humans carry out all these complex processes intuitively, instantaneously, and responsively.

Although spoken language is a naturally comfortable mode of communication for humans, ASR technology aims to harness this ease and transform it into a format beneficial for computers. It accomplishes this by converting speech signals into written text, thus transcribing oral communication into a computer-readable input.

Whereas oral communication is natural for humans, it presents significant challenges for computers due to its complexities. These complexities include physical aspects such as diverse accents, dialects, pitch, multiple speakers, and overlapping speech. Specific environmental factors can also add difficulties, such as channel distortions, background noises, and limitations in sampling and compression techniques. Additional complexities are associated with cognitive process aspects. Since spoken language carries inherent meaning, it's crucial for machines to understand words in their respective contexts to yield accurate transcriptions (e.g., distinguishing between \textit{knight} and \textit{night}). Further complexities arise during the transcription phase, which involves converting spoken words into written text, such as the need for appropriate punctuation.

Previous studies have addressed these challenges by using algorithmic approaches that leverage extensive datasets. For instance, \cite{christensen2001punctuation} introduced statistical prosody models for punctuation annotation, while \cite{wang2018supervised} surveyed models for speaker-noise separation. However, these solutions frequently encounter issues with generalization across various settings, including differences in speakers, languages, and environmental factors. Furthermore, these methodologies often face limitations in their accuracy and performance, which can hinder their effectiveness in diverse and complex real-world applications.

\subsection{ASR in the age of Large Language Models}
Large language models (LLMs; e.g., ChatGPT by OpenAI \cite{roumeliotis2023chatgpt}, Bard by Google \cite{manyika2023}) are revolutionizing the tools we use and also hold promise for the ASR task. Based on transformer-based architecture \cite{zhao2023survey} and vast datasets, LLMs are currently recognized as state-of-the-art (SOTA) for typical natural language processing (NLP) tasks (e.g., \cite{hendy2023good}, \cite{9400837}, \cite{luo2022critical}).

The remarkable success of LLMs in text-based tasks has sparked interest in the field of speech processing to develop solutions for the ASR deficits mentioned earlier, based on similar principles: transformer-based architecture \cite{latif2023transformers} and large corpora. Currently, ASR models based on these principles (e.g., Whisper \cite{radford2023robust}, SpeechT5 \cite{ao2021speecht5}) are considered SOTA for most ASR tasks.

To fully realize the significant benefits of ASR and its wide range of applications, large, high-quality datasets are essential. These datasets should include a wide vocabulary, diverse speakers, and a variety of topics. Since languages differ in multiple aspects (e.g., phonetics, syntax, and semantic structure), one may utilize datasets tailored to each specific language the ASR should support. At present, only a few languages have access to datasets of sufficient size and quality. Many languages suffer from a lack of resources, which prevents their speakers from maximizing the potential of existing ASR technology. In these languages, a lack of resources poses a major barrier to the effective adoption and utilization of ASR technology. In light of these challenges, it's crucial to support languages in the digital world for fostering linguistic diversity and inclusivity. Making more speech and transcribed speech data available for use could significantly benefit less-prevalent languages. Specifically, in this project, we aim to provide such support for Hebrew speakers.

\subsection{Processing Hebrew Speech}
The processing of Hebrew speech is challenged by the lack of available data. Some innovative monolingual datasets do exist in Hebrew (\cite{sharoni23_interspeech}) and there are some multilingual datasets that include Hebrew \cite{black2019cmu}. However, their scope and quality do not meet the requirements needed for training and optimizing robust ASR models. As a result, Hebrew speakers cannot fully benefit from advancements in ASR technology. Therefore, it is crucial to collect and develop comprehensive datasets in Hebrew.

\subsection{The Present Dataset}

The \textit{ivrit.ai} dataset that we present here has the potential to contribute substantially to various speech and text tasks in Hebrew. This dataset includes over \totalHours hours of Hebrew speech, collected from multiple online sources. The dataset includes a wide range of speakers, varying by gender, origin, and education level, as well as a variety of speech styles (e.g., official lectures, informal conversations), and topics (e.g., sports podcasts, Talmud lessons). Approximately 2.8 million utterances and about \totalWords words are included in the dataset. The audio data and transcripts are available for research and development in AI modeling, encompassing both non-commercial and commercial uses.

\section{Related works} 
Significant advances in ASR and NLP have been made in recent years, as a result of numerous datasets and research studies. However, the Hebrew language has received very little attention in this domain.

\subsection{General Speech Datasets and Research} 
Monolingual speech datasets have been developed for specific domains and communication situations, including the ATIS corpus for air travel information requests \cite{hemphill1990atis}, as well as other corpora containing recordings of meetings \cite{garofolo2004nist}, telephone conversations \cite{canavan1997callhome}, and broadcast media \cite{garofolo20042002}. The original TIMIT collection \cite{garofolo1993timit} and many of the broadcast news corpora provide relatively clean audio scenarios in which a single speaker reads from a prepared text, such as TREC \cite{garofolo2000trec} and CLEF \cite{federico2004clef}. This characteristic restricts the usefulness of the data in more dynamic environments and in more spontaneous situations. There are also collections of more naturally occurring conversational material, such as the CALLHOME corpus \cite{canavan1997callhome}, the Santa Barbara Corpus of Spoken American English \cite{du2000santa}, and the TED talks corpus \cite{hasebe2015design}, as well as podcast corpora, for instance, Spotify \cite{clifton2020spotify} and MSP \cite{lotfian2017building} \cite{martinez2020msp}. These collections capture unscripted and spontaneously organized discourse in a conversational setting, including turns, interviews, stretches of monologue, and argumentation. Due to this aspect, this data is more useful for dealing with natural language patterns and real-world dialogue dynamics. None of these datasets include Hebrew speech.

\subsection{Multilingual Speech Datasets and Research} 
Current multilingual speech datasets with transcriptions, such as those drawn from conversational telephone speech (IARPA Babel Program \cite{cui2013developing}), political speech \cite{wang2021voxpopuli}, and audiobooks \cite{panayotov2015librispeech} \cite{pratap2020mls}, cover a variety of domains and span approximately 100 languages. However, the representation of Hebrew in these datasets is relatively low. This limitation also extends to multilingual datasets without transcriptions, such as VoxLingua107 \cite{valk2021voxlingua107} and VoxPopuli \cite{wang2021voxpopuli}, which span multiple languages and contain large amounts of unlabeled data. Even in datasets like the one utilizing read versions of the New Testament (MSU \cite{black2019cmu} and recently MMS \cite{pratap2023scaling}), which cover many languages and provide high-quality alignments, the Hebrew content is not extensive. Unfortunately, the authors don't detail the Hebrew content, but generally suggest around 25 hours per language. The data from these datasets is used to train self-supervised models, build speech recognition systems, and develop language identification models \cite{pratap2023scaling} \cite{wang2021voxpopuli}. Despite the availability of numerous monolingual and multilingual speech datasets, each with their unique characteristics, limitations, and areas of focus, the representation of Hebrew remains limited. The field continues to evolve with ongoing efforts to develop more comprehensive and diverse datasets that can support a wider range of ASR and NLP research. Thus, there is a clear need for more extensive Hebrew content in these resources.

\subsection{Prior Hebrew Speech Datasets}
\begin{table*}[h]
\centering
\renewcommand{\arraystretch}{1.2}  
\small  
\begin{tabular}{|p{3cm}|l|l|l|l|p{3cm}|p{3cm}|}
\hline
Corpus & Hours & Speakers & Trans. & Type & Topics & License \\
\hline
SASPEECH \cite{sharoni23_interspeech} & $30$ & 1 & 4M/26A & Mixed & Economy, Politics & Non-commercial \\
HUJI Corpus \cite{marmorstein2022huji}& 3.8 & 60 & M & Conversations  & General Lifestyle & CC BY 4.0\\
CoSIH \cite{izre2001designing}& 12.3 & $\pm140$ & M & Conversations & General Lifestyle & Non-commercial \\
MaTaCOp \cite{azogui2016open}& 5.3 & 16 & M & Conversations & Map Task framework \cite{anderson1991hcrc} & Non-commercial\\
MLS \cite{pratap2020mls}& 126 & $13$ & M & Reading & Classic Books & CC BY 4.0 \\
CMU \cite{black2019cmu} & $\pm25$ & - & M & Reading & New Testament & -\\
MMS \cite{pratap2023scaling}& $\pm25$ & - & M & Reading & New Testament & -\\
Whisper \cite{radford2023robust}  & $688$ & - & - & - & - & Not available\\
ivrit.ai  & \totalHours & $+1000$ & A & Mixed & Wide range (economy, politics, science, bible, philosophy, technology, history, etc.) & Augmented CC BY 4.0 (see \nameref{subsec:availability} section) \\
\hline
\end{tabular}
\caption{Comparison of various speech datasets. The columns show the name of the corpus, total hours of audio available, the number of speakers included, whether the dataset is transcribed or not (M for manual transcription, and A for automatic transcription, the type of speech included (reading, conversations, mixed), the topics covered in the dataset, and the terms of the license for using the dataset. Dash (-) indicates that the data is not available for the corresponding field.}
\label{table:1}
\end{table*}

An array of datasets spanning academic to non-academic ventures provide resources for Hebrew ASR research. These datasets can be monolingual, containing Hebrew exclusively, or multilingual, encompassing Hebrew along with other languages. Table~\ref{table:1} furnishes a detailed overview of these available Hebrew speech datasets, both mono- and multilingual, outlining their distinct characteristics and inherent limitations.

\subsection{Natural Language Processing and Speech Recognition Research in Hebrew}
Substantial progress has been made in the field of Hebrew NLP, with models like Hebert \cite{chriqui2022hebert}, AlephBERT \cite{seker2021alephbert}, and AlephBERTGimmel \cite{guetta2022large} setting benchmarks in various textual tasks. These advancements illustrate the potential for an intricate understanding of written Hebrew texts.

The current landscape of Hebrew Automatic Speech Recognition (ASR) is largely supported by commercial ASR services like Google Cloud, Microsoft (e.g., via Word), Samsung, IBM Watson, and WIT.ai. These services offer engines with accessible APIs that support Hebrew \cite{silber2021cross}. Additionally, openly accessible models like OpenAI's Whisper \cite{radford2023robust} and Meta AI's multilingual model \cite{pratap2023scaling} are available as open sources. This availability allows developers to create their own models for specific tasks.
Expanding on these resources, we have contributed to Hebrew ASR development by introducing the \textit{ivrit.ai} dataset. This open-access dataset, free for all interested parties, equips researchers and developers with the means to further improve their Hebrew ASR models.

\subsection{Unlabeled Speech Datasets and Research}
In the field of ASR, significant progress has been achieved through the application and evolution of unsupervised speech pre-training techniques \cite{wu2020self}, semi-supervised learning (self-training) \cite{liu2022audio}, and the combination of these techniques \cite{xu2021self}. However, the research community focuses primarily on ASR tasks utilizing English as the main input. Through their proficient use of a copious amount of unlabeled English speech data, these methods have improved English-centric ASR applications. A recently published unlabeled speech corpus, covering 23 languages, demonstrated the potential of these techniques for more languages \cite{wang2021voxpopuli}. However, Hebrew was not included in this corpus. These results underscore the urgent need for large datasets of Hebrew speech, even if they are not labeled. Significant progress can be expected in Hebrew ASR if these datasets are developed and utilized effectively.

\section{Dataset Creation}
\begin{figure}[h]
  \centering
\begin{tikzpicture}[node distance=3cm, every node/.style={text centered}]
    \node (download) {Download};
    \node (vad) [right of=download] {VAD Processing};
    \node (transcription) [right of=vad] {Transcription};

    \draw[->] (download) -- (vad);
    \draw[->] (vad) -- (transcription);

    \node (symbol1) [above of=download, yshift=-1.6cm] {\includegraphics[width=1.5cm]{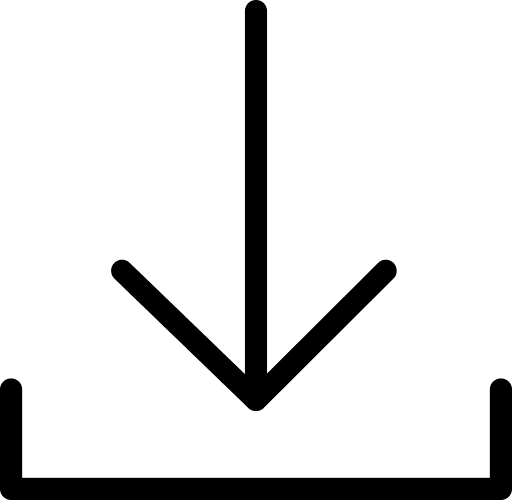}}; 
    \node (symbol2) [above of=vad, yshift=-1.6cm] {\includegraphics[width=1.5cm]{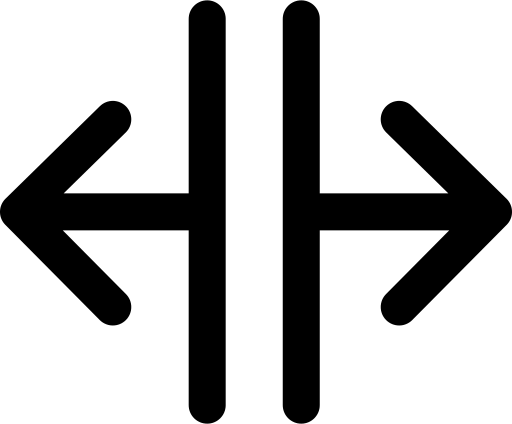}}; 
    \node (symbol3) [above of=transcription, yshift=-1.6cm] {\includegraphics[width=1.5cm]{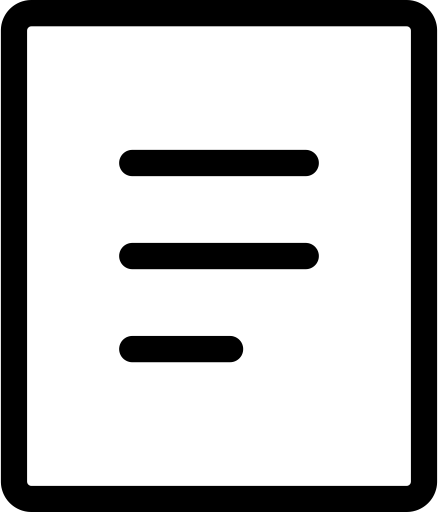}}; 
  \end{tikzpicture}
\caption{Illustration of the Data Pipeline: (1) Downloading the data in accordance with data creator agreements, (2) Processing Voice Activity Detection (VAD) to segment audio based on silences, and (3) Transcribing the segmented audio, which currently relies solely on machine-based process but is planned to incorporate human transcription.}
  \label{fig:pipeline}
\end{figure}
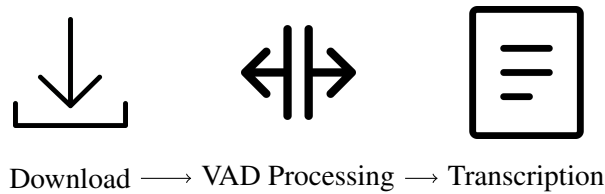

The data and code are openly available at Hugging Face (\url{https://huggingface.co/ivrit-ai}) and the GitHub repository (\url{https://github.com/yairl/ivrit.ai}). Throughout this section, we will describe the process of creating \textit{ivrit.ai}. This resource, with its variety of speakers and audio qualities, offers a comprehensive representation of the Hebrew language in multiple contexts. Detailed information on how it was collected and preprocessed will be provided. Figure \ref{fig:pipeline} shows a schematic diagram of the dataset creation pipeline.

\subsection{Data Acquisition}
We gathered audio clips from a variety of sources, encompassing both individual and institutional content creators.

ivrit.ai's license is specifically designed to enable commercial use of this corpus for training AI models, such as speech-to-text or LLM models, while preserving the intellectual property right of the content owner. Every item in the corpus has been provided by a content owner who signed permission to use this item under ivrit.ai's license, thereby permitting the use of their work in this manner.

The advantages of such an agreement are twofold. First, it ensures fairness towards the content creators by explicitly stating in the agreement that their work can be used. Second, it provides certainty for research and development entities by confirming that the data has been collected with permission and is distributed within a suitable legal framework.

\subsection{Data Processing}

\begin{figure}[ht]
\centering
\includegraphics[width=9cm]{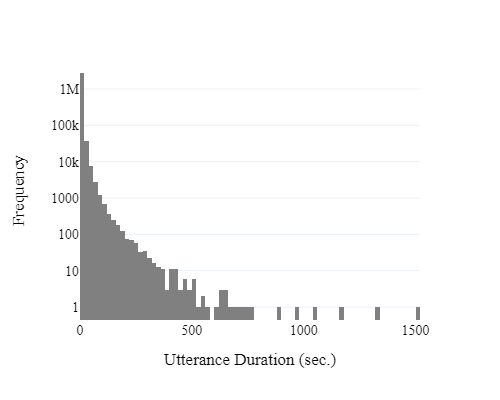}
\caption{Distribution of post-VAD audio clips in \textit{ivrit.ai} corpus. The x-axis represents the segment length  in seconds, while the y-axis indicates the frequency of each duration range}
\label{fig:utterance_duration}
\end{figure}

The \textit{ivrit.ai} data we've collected are being released in three distinct datasets, catering to various research needs:
\begin{itemize}

\item Raw Data: This is the original, unprocessed collection of audio clips.
\item Data Post-VAD: For this dataset, we have run Voice Activity Detection (VAD) on the raw audio \cite{SileroVAD}. This operation segregates short units, ranging from a few seconds to a minute, pinpointing parts where speakers were actively involved. Figure \ref{fig:utterance_duration} provides insights into the length distribution of the audio pieces post-VAD.
\item Partially Transcribed Data: This dataset provides a portion of the audio data along with their corresponding transcriptions.
\end{itemize}

\subsection{Transcribed Speech}
The \textit{ivrit.ai} dataset has been transcribed using the Whisper ASR tool (\cite{radford2023robust}, using the \textit{whisper-small} model). The transcription process is applied to numerous short audio segments, resulting in 2.8 million transcribed utterances and about \totalWords words. 

\section{Dataset Description}
\begin{figure}[ht]
\centering
\includegraphics[width=9cm]{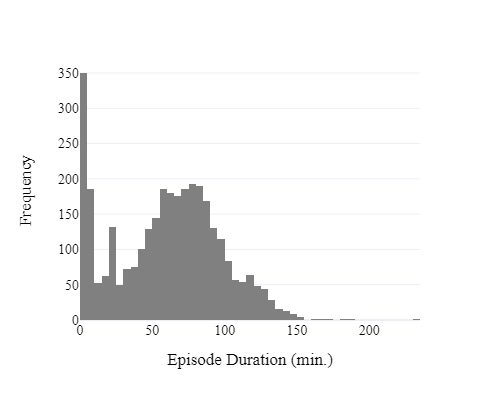}
\caption{Histogram depicting the distribution of episode durations in \textit{ivrit.ai} corpus. The x-axis represents the episode duration in minutes, while the y-axis indicates the frequency of each duration range}
\label{fig:episode_duration}
\end{figure}

As the data was collected from multiple contributors and represents a wide range of recordings (narrative podcast, conversation, lesson), we are not able to provide any precise information regarding the speakers. However, we can provide some general information.
The \textit{ivrit.ai} dataset comprises over \totalHours hours of speech from a thousand diverse speakers. The dataset encompasses a wide range of audio file lengths, and Figure \ref{fig:episode_duration} provides insights into the distribution of episode lengths within the dataset.

Speakers' ages ranged between the 20s and 70s. While some speakers are native Hebrew speakers, for others, Hebrew is the second language (with English, Russian, Arabic, and other languages being their native languages).

Languages other than Hebrew, predominantly English, can appear within the corpus at three levels of magnitude. First, as single words borrowed for specific uses, technical or otherwise; second, as slightly longer phrases that have gained popularity in Hebrew usage (for instance, the phrase "having said that" has been adopted into Hebrew conversations as is); and third, there are entire episodes that are conducted solely in English.

\section{Availability}
\label{subsec:availability}
The dataset is publicly accessible on the \textit{ivrit.ai} website and is distributed under an \textit{ivrit.ai} license, an augmented CC-BY 4.0 license tailored to allow AI model training for commercial use. Detailed licensing terms can be found in Appendix \ref{sec:appendix}.

\section{Discussion}

We present here the \textit{ivrit.ai} dataset, a comprehensive collection of over \totalHours hours of high-quality Hebrew speech, curated to advance AI research in the Hebrew language. This novel dataset consists of a wide range of speech types and topics, ready for use in various applications such as emergency response systems, accessibility tools for the disabled, medical transcription services, and digital voice assistants in the service industry, among others.

The \textit{ivrit.ai} dataset stands out among other Hebrew datasets in its size, diversity, and coverage of different speech styles and domains. Furthermore, the \textit{ivrit.ai} dataset offers more legal accessibility than many other datasets, as it is available for industrial use, making it a valuable resource for researchers and developers. Researchers can leverage this dataset to train and evaluate their models, while also utilizing it as a benchmark for performance comparison.

Among the limitations of the project, the dataset may be biased in aspects such as gender or age imbalances, which may affect the performance of AI models trained on the dataset. Additionally, the dataset's diversity, considered above as a strength, introduces variability in recording means, speaker characteristics, and background noises. Moreover, deficits in data collection and transcription could impact the dataset's quality or usability.

\section{Conclusion and Future Work}
ASR technology holds vast potential for enhancing various human processes. Although generating high-quality, efficient ASR tools is well known, model quality depends on the dataset size. Despite the benefits that can be obtained from ASR tools for some languages, others, such as Hebrew, are underwhelmed by the technology.

We introduced the \textit{ivrit.ai} dataset, a comprehensive collection of over \totalHours hours of Hebrew speech, designed to advance AI research in Hebrew. With a wide range of speech types and topics, the dataset offers many possibilities. In our view, the availability of such a diverse and extensive dataset is a significant step forward in the field of Hebrew ASR and NLP research. This dataset has the potential to improve multiple ASR-based systems' accuracy and performance.

Dataset acquisition tends to be effort-intensive, and fraught with legal difficulties due to copyright requirements that often conflict with standard licenses. \textit{ivrit.ai} aims to create the world's largest freely-available audio dataset in Hebrew, fully transcribed, and fully available for the specific purpose of training ASR and AI models.

 Looking forward, we plan to further expand the \textit{ivrit.ai} dataset, increase the corpus by another order of magnitude and promote applied developments based on the dataset, particularly in specific domains. Community involvement and collaboration will be crucial to these efforts. By making the dataset widely accessible, the aim is to place Hebrew at the forefront of AI research and technology.

\section{Acknowledgments}

We would like to express our deepest gratitude to all the content creators who generously allowed us to use their data for this project. Their contributions have been invaluable in advancing AI research in Hebrew. The full list of data contributors is updated and available on the \textit{ivrit.ai} website.

We also extend our heartfelt thanks to Adv. Eli Greenbaum from Yigal Arnon \& Co., who generously provided his legal expertise pro bono to draft the license for this open data project. His contribution has been instrumental in ensuring the accessibility and wide distribution of the \textit{ivrit.ai} dataset.

Your collective support and contributions have been instrumental in the success of this project, and we look forward to seeing the advancements in AI research that the \textit{ivrit.ai} dataset will facilitate.

\bibliography{anthology,custom}

\begin{thebibliography}{43}
\expandafter\ifx\csname natexlab\endcsname\relax\def\natexlab#1{#1}\fi

\bibitem[{Alarcon et~al.(2021)Alarcon, Moreno, and Martínez}]{9400837}
Rodrigo Alarcon, Lourdes Moreno, and Paloma Martínez. 2021.
\newblock \href {https://doi.org/10.1109/ACCESS.2021.3072697} {Lexical
  simplification system to improve web accessibility}.
\newblock \emph{IEEE Access}, 9:58755--58767.

\bibitem[{Anderson et~al.(1991)Anderson, Bader, Bard, Boyle, Doherty, Garrod,
  Isard, Kowtko, McAllister, Miller et~al.}]{anderson1991hcrc}
Anne~H Anderson, Miles Bader, Ellen~Gurman Bard, Elizabeth Boyle, Gwyneth
  Doherty, Simon Garrod, Stephen Isard, Jacqueline Kowtko, Jan McAllister, Jim
  Miller, et~al. 1991.
\newblock The hcrc map task corpus.
\newblock \emph{Language and speech}, 34(4):351--366.

\bibitem[{Ao et~al.(2021)Ao, Wang, Zhou, Wang, Ren, Wu, Liu, Ko, Li, Zhang
  et~al.}]{ao2021speecht5}
Junyi Ao, Rui Wang, Long Zhou, Chengyi Wang, Shuo Ren, Yu~Wu, Shujie Liu, Tom
  Ko, Qing Li, Yu~Zhang, et~al. 2021.
\newblock Speecht5: Unified-modal encoder-decoder pre-training for spoken
  language processing.
\newblock \emph{arXiv preprint arXiv:2110.07205}.

\bibitem[{Azogui et~al.(2016)Azogui, Lerner, and Silber-Varod}]{azogui2016open}
Jacob Azogui, Anat Lerner, and Vered Silber-Varod. 2016.
\newblock The open university of israel map task corpus (matacop).

\bibitem[{Black(2019)}]{black2019cmu}
Alan~W Black. 2019.
\newblock Cmu wilderness multilingual speech dataset.
\newblock In \emph{ICASSP 2019-2019 IEEE International Conference on Acoustics,
  Speech and Signal Processing (ICASSP)}, pages 5971--5975. IEEE.

\bibitem[{Canavan et~al.(1997)Canavan, Graff, and
  Zipperlen}]{canavan1997callhome}
Alexandra Canavan, David Graff, and George Zipperlen. 1997.
\newblock Callhome american english speech.
\newblock \emph{Linguistic Data Consortium}.

\bibitem[{Chriqui and Yahav(2022)}]{chriqui2022hebert}
Avihay Chriqui and Inbal Yahav. 2022.
\newblock Hebert and hebemo: A hebrew bert model and a tool for polarity
  analysis and emotion recognition.
\newblock \emph{INFORMS Journal on Data Science}, 1(1):81--95.

\bibitem[{Christensen et~al.(2001)Christensen, Gotoh, and
  Renals}]{christensen2001punctuation}
Heidi Christensen, Yoshihiko Gotoh, and Steve Renals. 2001.
\newblock Punctuation annotation using statistical prosody models.

\bibitem[{Clifton et~al.(2020)Clifton, Pappu, Reddy, Yu, Karlgren, Carterette,
  and Jones}]{clifton2020spotify}
Ann Clifton, Aasish Pappu, Sravana Reddy, Yongze Yu, Jussi Karlgren, Ben
  Carterette, and Rosie Jones. 2020.
\newblock The spotify podcast dataset.
\newblock \emph{arXiv preprint arXiv:2004.04270}.

\bibitem[{Cui et~al.(2013)Cui, Cui, Ramabhadran, Kim, Kingsbury, Mamou, Mangu,
  Picheny, Sainath, and Sethy}]{cui2013developing}
Jia Cui, Xiaodong Cui, Bhuvana Ramabhadran, Janice Kim, Brian Kingsbury,
  Jonathan Mamou, Lidia Mangu, Michael Picheny, Tara~N Sainath, and Abhinav
  Sethy. 2013.
\newblock Developing speech recognition systems for corpus indexing under the
  iarpa babel program.
\newblock In \emph{2013 IEEE International Conference on Acoustics, Speech and
  Signal Processing}, pages 6753--6757. IEEE.

\bibitem[{Du~Bois et~al.(2000)Du~Bois, Chafe, Meyer, Thompson, and
  Martey}]{du2000santa}
John~W Du~Bois, Wallace~L Chafe, Charles Meyer, Sandra~A Thompson, and Nii
  Martey. 2000.
\newblock Santa barbara corpus of spoken american english.
\newblock \emph{CD-ROM. Philadelphia: Linguistic Data Consortium}.

\bibitem[{Federico and Jones(2004)}]{federico2004clef}
Marcello Federico and Gareth~JF Jones. 2004.
\newblock The clef 2003 cross-language spoken document retrieval track.
\newblock In \emph{Comparative Evaluation of Multilingual Information Access
  Systems: 4th Workshop of the Cross-Language Evaluation Forum, CLEF 2003,
  Trondheim, Norway, August 21-22, 2003, Revised Selected Papers 4}, pages
  646--652. Springer.

\bibitem[{Garofolo et~al.(2004{\natexlab{a}})Garofolo, Fiscus, and
  Le}]{garofolo20042002}
J~Garofolo, Jonathan Fiscus, and Audrey Le. 2004{\natexlab{a}}.
\newblock 2002 rich transcription broadcast news and conversational telephone
  speech.
\newblock \emph{Linguistic Data Consortium}.

\bibitem[{Garofolo(1993)}]{garofolo1993timit}
John~S Garofolo. 1993.
\newblock Timit acoustic phonetic continuous speech corpus.
\newblock \emph{Linguistic Data Consortium, 1993}.

\bibitem[{Garofolo et~al.(2000)Garofolo, Auzanne, Voorhees
  et~al.}]{garofolo2000trec}
John~S Garofolo, Cedric~GP Auzanne, Ellen~M Voorhees, et~al. 2000.
\newblock The trec spoken document retrieval track: A success story.
\newblock \emph{NIST SPECIAL PUBLICATION SP}, 500(246):107--130.

\bibitem[{Garofolo et~al.(2004{\natexlab{b}})Garofolo, Laprun, Michel,
  Stanford, and Tabassi}]{garofolo2004nist}
John~S Garofolo, Christophe Laprun, Martial Michel, Vincent~M Stanford, and
  Elham Tabassi. 2004{\natexlab{b}}.
\newblock The nist meeting room pilot corpus.
\newblock In \emph{LREC}.

\bibitem[{Guetta et~al.(2022)Guetta, Shmidman, Shmidman, Shmidman, Guedalia,
  Koppel, Bareket, Seker, and Tsarfaty}]{guetta2022large}
Eylon Guetta, Avi Shmidman, Shaltiel Shmidman, Cheyn~Shmuel Shmidman, Joshua
  Guedalia, Moshe Koppel, Dan Bareket, Amit Seker, and Reut Tsarfaty. 2022.
\newblock Large pre-trained models with extra-large vocabularies: A contrastive
  analysis of hebrew bert models and a new one to outperform them all.
\newblock \emph{arXiv preprint arXiv:2211.15199}.

\bibitem[{Hasebe(2015)}]{hasebe2015design}
Yoichiro Hasebe. 2015.
\newblock Design and implementation of an online corpus of presentation
  transcripts of ted talks.
\newblock \emph{Procedia-Social and Behavioral Sciences}, 198:174--182.

\bibitem[{Hemphill et~al.(1990)Hemphill, Godfrey, and
  Doddington}]{hemphill1990atis}
Charles~T Hemphill, John~J Godfrey, and George~R Doddington. 1990.
\newblock The atis spoken language systems pilot corpus.
\newblock In \emph{Speech and Natural Language: Proceedings of a Workshop Held
  at Hidden Valley, Pennsylvania, June 24-27, 1990}.

\bibitem[{Hendy et~al.(2023)Hendy, Abdelrehim, Sharaf, Raunak, Gabr,
  Matsushita, Kim, Afify, and Awadalla}]{hendy2023good}
Amr Hendy, Mohamed Abdelrehim, Amr Sharaf, Vikas Raunak, Mohamed Gabr, Hitokazu
  Matsushita, Young~Jin Kim, Mohamed Afify, and Hany~Hassan Awadalla. 2023.
\newblock How good are gpt models at machine translation? a comprehensive
  evaluation.
\newblock \emph{arXiv preprint arXiv:2302.09210}.

\bibitem[{Izre'el et~al.(2001)Izre'el, Hary, and Rahav}]{izre2001designing}
Shlomo Izre'el, Benjamin Hary, and Giora Rahav. 2001.
\newblock Designing cosih: the corpus of spoken israeli hebrew.
\newblock \emph{International Journal of Corpus Linguistics}, 6(2):171--197.

\bibitem[{Latif et~al.(2023)Latif, Zaidi, Cuayahuitl, Shamshad, Shoukat, and
  Qadir}]{latif2023transformers}
Siddique Latif, Aun Zaidi, Heriberto Cuayahuitl, Fahad Shamshad, Moazzam
  Shoukat, and Junaid Qadir. 2023.
\newblock Transformers in speech processing: A survey.
\newblock \emph{arXiv preprint arXiv:2303.11607}.

\bibitem[{Liu et~al.(2022)Liu, Mallol-Ragolta, Parada-Cabaleiro, Qian, Jing,
  Kathan, Hu, and Schuller}]{liu2022audio}
Shuo Liu, Adria Mallol-Ragolta, Emilia Parada-Cabaleiro, Kun Qian, Xin Jing,
  Alexander Kathan, Bin Hu, and Bjoern~W Schuller. 2022.
\newblock Audio self-supervised learning: A survey.
\newblock \emph{Patterns}, 3(12).

\bibitem[{Lotfian and Busso(2017)}]{lotfian2017building}
Reza Lotfian and Carlos Busso. 2017.
\newblock Building naturalistic emotionally balanced speech corpus by
  retrieving emotional speech from existing podcast recordings.
\newblock \emph{IEEE Transactions on Affective Computing}, 10(4):471--483.

\bibitem[{Luo et~al.(2022)Luo, Lau, Li, and Si}]{luo2022critical}
Bei Luo, Raymond~YK Lau, Chunping Li, and Yain-Whar Si. 2022.
\newblock A critical review of state-of-the-art chatbot designs and
  applications.
\newblock \emph{Wiley Interdisciplinary Reviews: Data Mining and Knowledge
  Discovery}, 12(1):e1434.

\bibitem[{Manyika(2023)}]{manyika2023}
James Manyika. 2023.
\newblock \href {https://ai.google/static/documents/google-about-bard.pdf} {An
  overview of bard: an early experiment with generative ai}.

\bibitem[{Marmorstein and Matalon(2022)}]{marmorstein2022huji}
Michal Marmorstein and Nadav Matalon. 2022.
\newblock The huji corpus of spoken hebrew: An interaction-oriented design of a
  corpus.

\bibitem[{Martinez-Lucas et~al.(2020)Martinez-Lucas, Abdelwahab, and
  Busso}]{martinez2020msp}
Luz Martinez-Lucas, Mohammed Abdelwahab, and Carlos Busso. 2020.
\newblock The msp-conversation corpus.
\newblock \emph{Interspeech 2020}.

\bibitem[{Panayotov et~al.(2015)Panayotov, Chen, Povey, and
  Khudanpur}]{panayotov2015librispeech}
Vassil Panayotov, Guoguo Chen, Daniel Povey, and Sanjeev Khudanpur. 2015.
\newblock Librispeech: an asr corpus based on public domain audio books.
\newblock In \emph{2015 IEEE international conference on acoustics, speech and
  signal processing (ICASSP)}, pages 5206--5210. IEEE.

\bibitem[{Pratap et~al.(2023)Pratap, Tjandra, Shi, Tomasello, Babu, Kundu,
  Elkahky, Ni, Vyas, Fazel-Zarandi et~al.}]{pratap2023scaling}
Vineel Pratap, Andros Tjandra, Bowen Shi, Paden Tomasello, Arun Babu, Sayani
  Kundu, Ali Elkahky, Zhaoheng Ni, Apoorv Vyas, Maryam Fazel-Zarandi, et~al.
  2023.
\newblock Scaling speech technology to 1,000+ languages.
\newblock \emph{arXiv preprint arXiv:2305.13516}.

\bibitem[{Pratap et~al.(2020)Pratap, Xu, Sriram, Synnaeve, and
  Collobert}]{pratap2020mls}
Vineel Pratap, Qiantong Xu, Anuroop Sriram, Gabriel Synnaeve, and Ronan
  Collobert. 2020.
\newblock Mls: A large-scale multilingual dataset for speech research.
\newblock \emph{arXiv preprint arXiv:2012.03411}.

\bibitem[{Radford et~al.(2023)Radford, Kim, Xu, Brockman, McLeavey, and
  Sutskever}]{radford2023robust}
Alec Radford, Jong~Wook Kim, Tao Xu, Greg Brockman, Christine McLeavey, and
  Ilya Sutskever. 2023.
\newblock Robust speech recognition via large-scale weak supervision.
\newblock In \emph{International Conference on Machine Learning}, pages
  28492--28518. PMLR.

\bibitem[{Roumeliotis and Tselikas(2023)}]{roumeliotis2023chatgpt}
Konstantinos~I Roumeliotis and Nikolaos~D Tselikas. 2023.
\newblock Chatgpt and open-ai models: A preliminary review.
\newblock \emph{Future Internet}, 15(6):192.

\bibitem[{Seker et~al.(2021)Seker, Bandel, Bareket, Brusilovsky, Greenfeld, and
  Tsarfaty}]{seker2021alephbert}
Amit Seker, Elron Bandel, Dan Bareket, Idan Brusilovsky, Refael~Shaked
  Greenfeld, and Reut Tsarfaty. 2021.
\newblock Alephbert: A hebrew large pre-trained language model to start-off
  your hebrew nlp application with.
\newblock \emph{arXiv preprint arXiv:2104.04052}.

\bibitem[{Sharoni et~al.(2023)Sharoni, Shenberg, and
  Cooper}]{sharoni23_interspeech}
Orian Sharoni, Roee Shenberg, and Erica Cooper. 2023.
\newblock Saspeech: A hebrew single speaker dataset for text to speech and
  voice conversion.
\newblock In \emph{Proc. Interspeech 2023}, page To Appear.

\bibitem[{Silber-Varod et~al.(2021)Silber-Varod, Siegert, Jokish, Sinha, and
  Geri}]{silber2021cross}
Vered Silber-Varod, Ingo Siegert, Oliver Jokish, Yamini Sinha, and Nitza Geri.
  2021.
\newblock A cross-language study of speech recognition systems for english,
  german, and hebrew.
\newblock \emph{Online Journal of Applied Knowledge Management}, 9(1):1--15.

\bibitem[{Team(2021)}]{SileroVAD}
Silero Team. 2021.
\newblock Silero vad: pre-trained enterprise-grade voice activity detector
  (vad), number detector and language classifier.
\newblock \url{https://github.com/snakers4/silero-vad}.

\bibitem[{Valk and Alum{\"a}e(2021)}]{valk2021voxlingua107}
J{\"o}rgen Valk and Tanel Alum{\"a}e. 2021.
\newblock Voxlingua107: a dataset for spoken language recognition.
\newblock In \emph{2021 IEEE Spoken Language Technology Workshop (SLT)}, pages
  652--658. IEEE.

\bibitem[{Wang et~al.(2021)Wang, Riviere, Lee, Wu, Talnikar, Haziza,
  Williamson, Pino, and Dupoux}]{wang2021voxpopuli}
Changhan Wang, Morgane Riviere, Ann Lee, Anne Wu, Chaitanya Talnikar, Daniel
  Haziza, Mary Williamson, Juan Pino, and Emmanuel Dupoux. 2021.
\newblock Voxpopuli: A large-scale multilingual speech corpus for
  representation learning, semi-supervised learning and interpretation.
\newblock \emph{arXiv preprint arXiv:2101.00390}.

\bibitem[{Wang and Chen(2018)}]{wang2018supervised}
DeLiang Wang and Jitong Chen. 2018.
\newblock Supervised speech separation based on deep learning: An overview.
\newblock \emph{IEEE/ACM Transactions on Audio, Speech, and Language
  Processing}, 26(10):1702--1726.

\bibitem[{Wu et~al.(2020)Wu, Wang, Pino, and Gu}]{wu2020self}
Anne Wu, Changhan Wang, Juan Pino, and Jiatao Gu. 2020.
\newblock Self-supervised representations improve end-to-end speech
  translation.
\newblock \emph{arXiv preprint arXiv:2006.12124}.

\bibitem[{Xu et~al.(2021)Xu, Baevski, Likhomanenko, Tomasello, Conneau,
  Collobert, Synnaeve, and Auli}]{xu2021self}
Qiantong Xu, Alexei Baevski, Tatiana Likhomanenko, Paden Tomasello, Alexis
  Conneau, Ronan Collobert, Gabriel Synnaeve, and Michael Auli. 2021.
\newblock Self-training and pre-training are complementary for speech
  recognition.
\newblock In \emph{ICASSP 2021-2021 IEEE International Conference on Acoustics,
  Speech and Signal Processing (ICASSP)}, pages 3030--3034. IEEE.

\bibitem[{Zhao et~al.(2023)Zhao, Zhou, Li, Tang, Wang, Hou, Min, Zhang, Zhang,
  Dong et~al.}]{zhao2023survey}
Wayne~Xin Zhao, Kun Zhou, Junyi Li, Tianyi Tang, Xiaolei Wang, Yupeng Hou,
  Yingqian Min, Beichen Zhang, Junjie Zhang, Zican Dong, et~al. 2023.
\newblock A survey of large language models.
\newblock \emph{arXiv preprint arXiv:2303.18223}.

\end{thebibliography}
\bibliographystyle{acl_natbib}

\appendix

\section{License Appendix}
\label{sec:appendix}
This material and data is licensed under the terms of the Creative Commons Attribution 4.0 International License (CC BY 4.0), The full text of the CC-BY 4.0 license is available at https://creativecommons.org/licenses/by/4.0/.

Notwithstanding the foregoing, this material and data may only be used, modified and distributed for the express purpose of training AI models, and subject to the foregoing restriction. In addition, this material and data may not be used in order to create audiovisual material that simulates the voice or likeness of the specific individuals appearing or speaking in such materials and data (a “deep-fake”). To the extent this paragraph is inconsistent with the CC-BY-4.0 license, the terms of this paragraph shall govern.

By downloading or using any of this material or data, you agree that the Project makes no representations or warranties in respect of the data, and shall have no liability in respect thereof. These disclaimers and limitations are in addition to any disclaimers and limitations set forth in the CC-BY-4.0 license itself. You understand that the project is only able to make available the materials and data pursuant to these disclaimers and limitations, and without such disclaimers and limitations the project would not be able to make available the materials and data for your use.
\end{document}